\documentclass[aps,prl,preprintnumbers,amsmath,amssymb,twocolumn,superscriptaddress,showpacs,floatfix]{revtex4}
\usepackage{bbm}
\usepackage{mathrsfs}
\usepackage{amsmath,amssymb,bm,comment}
\usepackage{hyperref}
\usepackage{graphicx}
\usepackage{dcolumn} 
\usepackage{bm} 

\begin{document}

\begin{flushright}
{\mbox{\hspace{10cm}}} USTC-ICTS-12-02
\end{flushright}

\title{Chiral anomaly and local polarization effect from quantum kinetic approach}

\author{Jian-Hua Gao}
\affiliation{School of Space Science and Physics, Shandong
University at Weihai, Weihai 264209, China}

\affiliation{Interdisciplinary Center for Theoretical Study and
Department of Modern Physics, University of Science and Technology
of China, Hefei 230026, China}

\author{Zuo-Tang Liang}
\affiliation{School of Physics, Shandong University, Jinan, Shandong 250100, China}

\author{Shi Pu}
\affiliation{Interdisciplinary Center for Theoretical Study and
Department of Modern Physics, University of Science and Technology
of China, Hefei 230026, China}

\author{Qun Wang}
\affiliation{Interdisciplinary Center for Theoretical Study and
Department of Modern Physics, University of Science and Technology
of China, Hefei 230026, China}

\author{Xin-Nian Wang}
\affiliation{Key Laboratory of Quark and Lepton Physics (MOE) and Institute of Particle Physics, Central China Normal University, Wuhan, 430079, China}
\affiliation{Nuclear Science Division, MS 70R0319, Lawrence Berkeley National Laboratory,
Berkeley, California 94720}

\date{\today}

\begin{abstract}

A power expansion scheme is set up to determine  the Wigner function 
that satisfies the  quantum kinetic equation for spin-1/2 charged fermions in a background
electromagnetic field. Vector and axial-vector current induced by magnetic
field and vorticity are obtained simultaneously from the Wigner function.
The chiral magnetic and vortical effect and chiral anomaly are shown as
natural consequences of the quantum kinetic equation. The axial-vector current induced by 
vorticity is argued to lead to a local polarization
effect along the vorticity direction in heavy-ion collisions.

\end{abstract}


\maketitle


{\it  Introduction.} ---
Chiral anomaly is an important quantum effect which is absent at the
classical level. Recently it has been shown that such a microscopic
quantum effect can have a macroscopic impact on the dynamics of
relativistic fluids, termed as the chiral magnetic and vortical effect (CME and CVE)
\cite{Kharzeev:2007jp,Fukushima:2008xe,Kharzeev:2010gr} as manifested in
currents induced by magnetic field and vorticity. Such effects and related topics have
been investigated within a variety of approaches, such as AdS/CFT duality
\cite{Erdmenger:2008rm,Banerjee:2008th,Torabian:2009qk,Rebhan:2009vc,Kalaydzhyan:2011vx},
relativistic hydrodynamics
\cite{Son:2009tf,Pu:2010as,Kharzeev:2011ds}, and quantum field theory
\cite{Metlitski:2005pr,Newman:2005as,Fukushima:2008xe,
Charbonneau:2009ax,Lublinsky:2009wr,Asakawa:2010bu,Landsteiner:2011cp}.
However,  it is still not clear how CME and CVE can emerge from a microscopic quantum 
kinetic theory. 

In this Letter we make a first attempt to derive both the CME and CVE from a quantum kinetic theory.
 A power expansion in space-time derivatives and weak external fields is used to
determine the analytic form of vector and axial-vector components of the Wigner function that 
satisfies the quantum kinetic equation for spin-1/2 massless fermions. 
The CME and CVE appear naturally in the induced currents. Chiral anomaly and other 
conservation laws are also automatically satisfied.
The axial-vector current induced by vorticity depends
quadratically on the temperature, baryonic and chiral chemical
potential. So it should be present in both hot and
dense matter, and can lead to a local polarization effect in
heavy-ion collisions as proposed in earlier studies
\cite{Liang:2004ph,Becattini:2007sr,Gao:2007bc}. This
provides another possible future experimental measurement of the CVE in
high-energy heavy-ion collisions.

The quantum kinetic approach can provide a bridge between the microscopic and macroscopic
description of the CME and CVE and should be more suitable for future simulations of 
both effects in heavy-ion collisions. The power expansion method can also be applied to the calculation 
of other transport coefficients.


{\it Quantum kinetic equation.} ---
In a quantum kinetic theory, the classical phase-space distribution $f(x,p)$ is
replaced by the Wigner function $W(x,p)$ in space-time $x$ and
4-momentum $p$, defined as the ensemble average of
the Wigner operator \cite{Elze:1986qd,Vasak:1987um,Elze:1989un}
for spin-1/2 fermions,
\begin{equation}
\label{wigner}
\hat W_{\alpha\beta} = \int\frac{d^4 y}{(2\pi)^4}
e^{-ip\cdot y} \bar\psi_\beta(x_+)U(x_+,x_-) \psi_{\alpha}(x_-),
\end{equation}
where $\psi_{\alpha}$ and $\bar\psi_{\beta}$ are Dirac spinor fields, 
$x_{\pm}\equiv x \pm \frac{1}{2}y$ are two space-time points centered at
$x$ with space-time separation $y$, and the gauge link $U$,
\begin{equation}
\label{link}
U (x_+,x_- ) \equiv e^{-i Q \int_{x_-}^{x_+} dz^\mu A_\mu (z)},
\end{equation}
ensures the gauge invariance of $\hat W_{\alpha\beta}$. Here $Q$ is
the electromagnetic charge of the fermions, and $A_\mu$ is the electromagnetic vector potential. 
Note that we use the metric convention $g^{\mu\nu}=\mathrm{diag}(1,-1,-1,-1)$. To
simplify the quantum kinetic equation under a background field we
consider a massless and collisionless fermionic system in a constant
external electromagnetic field $F_{\mu\nu}$ in the lab frame. 
Since we only consider a classical
background field, we have dropped the path ordering in
the gauge link in Eq.~(\ref{link}). The Wigner function is a matrix
in Dirac space and satisfies the quantum kinetic equation
\cite{Elze:1986qd,Vasak:1987um,Elze:1989un},
\begin{equation}
\label{eq-c}
\gamma_\mu ( p^\mu +\frac{i}{2}\nabla^\mu)  W(x,p)=0 ,
\end{equation}
where $\gamma^{\mu}$'s are Dirac matrices and
$\nabla^\mu \equiv \partial^\mu_x - Q {F^\mu}_\nu\partial^\nu_p$.
The Wigner function should contain information about quantum interactions and we will prove
that all currents including chiral anomaly can be derived from the above equation.
To this end, we decompose the Wigner function in terms of 16
independent generators of the Clifford algebra,
\begin{eqnarray}
\label{decomposition}
W(x,p)&=&\left. \frac{1}{4}\right[\mathscr{F}(x,p)+i\gamma^5 \mathscr{P}(x,p)
+\gamma^\mu \mathscr{V}_\mu (x,p) \nonumber \\
&&\left. +\gamma^5 \gamma^\mu \mathscr{A}_\mu (x,p)
+\frac{1}{2}\sigma^{\mu\nu} \mathscr{S}_{\mu\nu}(x,p)\right].
\end{eqnarray}
Eq.~(\ref{eq-c}) then leads to two decoupled sets of equations \cite{Elze:1986qd,Vasak:1987um,Elze:1989un},
one of which relevant to our study reads,
\begin{eqnarray}
\label{V-c-A-c}
p^\mu\mathscr{V}_\mu &=&0,\ \  p^\mu\mathscr{A}_\mu = 0,\\
\label{V-eq-A-eq}
\nabla^\mu\mathscr{V}_\mu &=& 0,\ \ \nabla^\mu\mathscr{A}_\mu = 0,\\
\label{VA}
\epsilon_{\mu\nu\rho\sigma}\nabla^\rho \mathscr{A}^\sigma
&=& -2\left(p_\mu \mathscr{V}_\nu-p_\nu \mathscr{V}_\mu\right),\\
\label{AV}
\epsilon_{\mu\nu\rho\sigma}\nabla^\rho \mathscr{V}^\sigma
&=& -2\left(p_\mu \mathscr{A}_\nu-p_\nu \mathscr{A}_\mu\right),
\end{eqnarray}
where $\epsilon^{\mu\nu\rho\sigma}$ is the Levi-Civita
anti-symmetric tensor,
$\mathscr{V}_\mu (x,p)$ and $\mathscr{A}_\mu (x,p)$
are the vector and axial-vector component of the Wigner function, 
which will give rise to the vector and axial-vector current,
respectively, after integration over four-momentum.




{\it Power expansion.} ---
We assume a system close to local equilibrium under a constant external field $F^{\mu\nu}$. 
Therefore, $\mathscr{V}_\mu(x,p)$ and $\mathscr{A}_\mu(x,p)$ 
will depend on $x$ only through fluid four-velocity $u(x)$,
temperature $T(x)$, chemical potential $\mu(x)$ and chiral chemical potential $\mu_5(x)$. 
We will determine the analytic form of the Wigner function in terms of \{$p, F^{\mu\nu}, u, T, \mu, \mu_{5}$\}
from the kinetic equation.

We further assume that the space-time derivative $\partial_x$ and the field strength
$F_{\mu\nu}$ are small variables of the same order
and can be used as parameters in the power expansion 
of $\mathscr{V}_\mu$ and $\mathscr{A}_\mu$ (similar to the Knudsen number expansion in
hydrodynamics),
\begin{equation}
\mathscr{V}^\mu = \mathscr{V}^\mu_0 + \mathscr{V}^\mu_1 + \cdots \, ,\ \
\mathscr{A}^\mu = \mathscr{A}^\mu_0 + \mathscr{A}^\mu_1 + \cdots\, ,
\end{equation}
where the subscripts $0,1,...$ denote orders of the power expansion.
Note that $\mathscr{V}_{n}^\mu$ and $\mathscr{A}_{n}^\mu$ are
related to $\mathscr{A}_{n-1}^\mu$ and $\mathscr{V}_{n-1}^\mu$  via
Eqs.~(\ref{VA}-\ref{AV}) ($n\geq 1$). One can therefore
use an iterative scheme to solve $\mathscr{V}_\mu$ and
$\mathscr{A}_\mu$ order by order.

Note that the field strengths $F^{\mu\nu}$ are assumed to be constant in the lab frame.
Later, we have to define electromagnetic fields in the
local comoving frame of a fluid cell, $E_\sigma = u^\rho F_{\sigma\rho}$, $B_\sigma
=(1/2)\epsilon_{\sigma\mu\nu\rho}u^\mu F^{\nu\rho}$,  which depend on $x$ via the fluid 
velocity $u(x)$. The space-time derivative $\partial_x$ is then given by
\begin{equation}
\label{partial-x}
\partial _\sigma ^x = \partial _\sigma T \frac{\partial }{\partial T}
+ \partial _\sigma u_\rho \frac{\partial }{\partial u_\rho} +
\partial _\sigma \mu \frac{\partial }{\partial \mu } +
\partial _\sigma \mu _5 \frac{\partial }{\partial \mu _5}.
\end{equation}


{\it Zeroth-order Wigner function.} ---
In general,  $\mathscr{V}^\mu_0$ and $\mathscr{A}^\mu_0$ can only have two terms,
each proportional to the zeroth-order four-vectors $p^{\mu}$ or $u^{\mu}$ with a total
of four independent coefficients. Since the left-hand sides of Eqs.~(\ref{VA}-\ref{AV}) 
are at least of first order, the zeroth-order terms on the right-hand sides must vanish, 
which set the coefficients of the $u^{\mu}$-terms to be zero. With additional constraints by
Eq.~(\ref{V-c-A-c}), $\mathscr{V}^\mu_0$ and $\mathscr{A}^\mu_0$ have to take the following forms,
\begin{equation}
\label{Vmu-0-1}
\mathscr{V}^\mu_0 = p^\mu \delta\left(p^2\right)V_{0}, \ \
\mathscr{A}^\mu_0 = p^\mu \delta\left(p^2\right)A_{0},
\end{equation}
where $V_{0}$ and $A_{0}$ are the phase space distributions of
massless spin-1/2 fermions at the zeroth order and cannot 
be determined by Eqs.~(\ref{V-c-A-c}-\ref{AV}). We assume they 
take the equilibrium form, 
\begin{eqnarray}
\label{disV}
\left[V_{0},A_{0}\right] &=& 
\sum _{s=\pm 1} \theta (s u\cdot p) \left[(f_{s,R}+ f_{s,L}), (f_{s,R}- f_{s,L})\right], \nonumber \\
f_{s,\chi}& = &\frac{2}{(2\pi)^3}\frac{1}{e^{s(u\cdot p-\mu _\chi)/T}+1}, (\chi=R,L),
\end{eqnarray}
where $R (L)$ denotes the right (left)-handed fermions 
and $\mu _{R,L}=\mu \pm \mu_5$ \cite{Fukushima:2008xe}. Note
that $V_{0}$ ($A_{0}$) is the sum (difference) of two positive distributions
for any values of $\mu$ and $\mu_{5}$. This asymmetry between $V_{0}$ and $A_{0}$ as
inputs to the iterative operation will feed down to the first-order Wigner 
functions  $\mathcal{V}^{\mu}_{1}$ and $\mathcal{A}^{\mu}_{1}$ and the final vector 
and axial-vector currents, even though the kinetic equations in Eqs. (\ref{V-c-A-c}-\ref{AV}) are symmetric
for $\mathcal{V}^{\mu}$ and $\mathcal{A}^{\mu}$.


The zeroth-order Wigner functions should also satisfy Eq.~(\ref{V-eq-A-eq}), 
which provides constraints on fluid and thermodynamical variables. 
Substitute Eqs.~(\ref{Vmu-0-1}-\ref{disV}) into Eq.~(\ref{V-eq-A-eq}),
we obtain $\nabla_\mu \mathscr{V}_0^\mu$ and $\nabla_\mu \mathscr{A}_0^\mu$ as
sums of six independent terms involving the momentum vector
$\bar{p}_\sigma \equiv \Delta _{\sigma\rho } p^\rho $
($\Delta_{\sigma\rho }\equiv g_{\sigma\rho }-u_\sigma u_\rho $),
tensor $\bar{p}_\sigma\bar{p}_\rho$, scalars $\bar{p}^2$ and
$u\cdot p$. To ensure $\nabla_\mu \mathscr{V}_0^\mu =\nabla_\mu \mathscr{A}_0^\mu=0$
for any values of $p$, these six terms all have to vanish, resulting in the following 
constraints at the first order,
\begin{eqnarray}
\label{relation1}
&& \Delta^{\sigma\alpha}\Delta^{\rho\beta}\left(\partial_\alpha u_\beta + \partial_\beta u_\alpha
-\frac{2}{3}\Delta_{\alpha\beta}\Delta^{\rho\sigma}\partial_\rho u_\sigma\right)=0,\nonumber\\
\label{relation2}
&& T\Delta^{\sigma\rho}\partial_\rho \frac{\mu}{T} + QE^\sigma =0, \nonumber\\
&& u\cdot\partial u^\sigma -\Delta^{\sigma\rho }\partial_\rho \ln T=0, \nonumber\\
&& \partial_\sigma \frac{\mu_5}{T}=0,\;\;\;\; u^\sigma\partial_\sigma \frac{\mu}{T}=0, \nonumber\\
&& u\cdot\partial T +\frac{1}{3}T\Delta^{\rho\sigma}\partial_\rho
u_\sigma=0.
\end{eqnarray}
Note that we have dropped $\delta(p_0)$ terms
from derivatives of $\theta(p_0)$ and $\theta(-p_0)$, which
are irrelevant when carrying out the 4-momentum integration due to
vanishing phase space at zero momentum. 
Since we are interested in currents
induced by external fields and vorticity, we consider
only the static case with a constant temperature.
The above constraints  are reduced to,
\begin{eqnarray}
\label{relation-static}
&& u\cdot\partial u^\sigma =0,\; 
\partial _\sigma u^\sigma =0, \; \partial_\sigma \mu = - QE_\sigma , \nonumber \\
&& \mu_5=\mathrm{const},\;\;\;\; (\mathrm{for}\;\; T=\mathrm{const.})
\end{eqnarray}
which has a simple solution  $\mu={\rm const.}-QE\cdot x$ 
and a solenoidal fluid velocity $u^\sigma (\mathbf{x}-\mathbf{u}t/u^0)$ 
with ${\bf \partial}\cdot{\bf u} =0$.

{\it First-order Wigner function.} ---
With the zeroth-order Wigner functions in Eqs.~(\ref{Vmu-0-1}-\ref{disV}) one can determine the first-order
$\mathscr{V}_1^\mu$ and $\mathscr{A}_1^\mu$ from Eqs.~(\ref{V-c-A-c}-\ref{AV}).
A general form linear in the first-order variables $X^\mu = (E^\mu, B^\mu, \omega^\mu )$ and
constrained by Eq.~(\ref{V-c-A-c}) can be written as,
\begin{eqnarray}
\label{Vmu-1}
\mathscr{Z}_1^\mu&=&
\sum_{X=E,B,\omega } \left[ u_\nu (g^{\nu\mu}-p^\nu p^\mu/p^2)
p^2 (\bar p\cdot X ) Z_{X1} \right. \nonumber\\
&&+X_\nu (g^{\nu\mu}-p^\nu p^\mu/p^2) p^2 Z_{X2} \nonumber \\
&&+X_\nu (g^{\nu\mu}-\bar{p}^\nu \bar{p}^\mu/\bar{p}^2) \bar{p}^2 Z_{X3}\nonumber\\
&&\left. +\epsilon^{\mu\lambda\rho\sigma}u_\lambda p_\rho X_\sigma
Z_{X4} \right],
\end{eqnarray}
where $\mathscr{Z}_1^\mu=(\mathscr{V}_1^\mu,\mathscr{A}_1^\mu)$ and 
 $\omega_\mu = (1/2)\epsilon_{\mu\nu\rho\sigma}u^\nu \partial^\rho u^\sigma $
is the fluid vorticity. Note that $X\cdot u=0$. 
There are 24 independent coefficients $Z_{Xi}=(V_{X_i}, A_{Xi})$ $(i=1,2,3,4)$ 
in the above power expansion of the first order.  
With Eqs.~(\ref{Vmu-0-1}) and (\ref{Vmu-1}) for the zeroth and 
first-order Wigner functions, 
both of Eqs.~(\ref{VA}) and (\ref{AV}) at the first-order contain 3 different tensor
structures, each consisting of terms linear in three independent variables
$X^{\mu}=(E^{\mu},B^{\mu},\omega^{\mu})$. Setting these terms to vanish separately
gives 18 equations which leave only 6 of the 24 coefficients in Eq.~(\ref{Vmu-1}) undetermined.
Further requiring Eq.~(\ref{V-eq-A-eq}) be satisfied by $\mathscr{V}_1^\mu$ and $\mathscr{A}_1^\mu$, 
we can obtain the unique forms of $\mathscr{V}_\mu$ and $\mathscr{A}_\mu$ to the first order,
\begin{eqnarray}
\label{V-final}
\mathscr{Z}^\mu &=& p^\mu \delta (p^2)Z_{0}
+ \frac{1}{2} p_\nu [ u^\mu \omega^\nu - u^\nu \omega ^\mu ]
\frac{\partial \bar{Z_{0}}}{\partial (u\cdot p)}  \delta ( p^2 )\nonumber\\
& & - Q p_\nu [ u^\mu B^\nu - u^\nu B^\mu ] \bar{Z}_{0} \delta' ( p^2 )\nonumber\\
& & + Q \epsilon^{\mu\lambda\rho\sigma } u_\lambda p_\rho E_\sigma
\bar{Z}_{0} \delta' ( p^2 ) ,
\end{eqnarray}
where $\mathscr{Z}=(\mathscr{V}, \mathscr{A})$, $Z_{0}=(V_{0},A_{0})$ and $\bar{Z_{0}}=(A_{0},V_{0})$.


{\it Induced currents, CME and CVE.} ---
We can derive the vector and axial-vector current from the above Wigner
functions up to the first order in power expansion,
\begin{eqnarray}
\label{jjv}
j^\mu &=&\int d^4p \mathscr{V}^\mu = n u^\mu +\xi\omega^\mu +\xi_B B^\mu , \\
\label{jj5}
j_5^\mu &=& \int d^4p \mathscr{A}^\mu = n_5 u^\mu+\xi_5\omega^\mu + \xi_{B5} B^\mu .
\end{eqnarray}
The energy-momentum tensor $T^{\mu\nu}$ can also be evaluated,
\begin{eqnarray}
\label{emt}
T^{\mu\nu} &=& \frac 12 \int d^4p (p^\mu \mathscr{V}^\nu
+ p^\nu \mathscr{V}^\mu) \nonumber \\
&=& (\epsilon + P)u^{\mu}u^{\nu}-Pg^{\mu\nu}+ n_5 (u^\mu \omega^\nu + u^\nu \omega^\mu )
\nonumber\\
&& + \frac 12 Q \xi (u^\mu B^\nu + u^\nu B^\mu ).
\end{eqnarray}
The charge $n$, $n_5$  and energy density $\epsilon$ in equilibrium,
\begin{equation}
N_{0}= 2\pi \int d p_0 p_0^{i} [\theta (p_0)-\theta (-p_0)] Z_{N0} ,
\end{equation}
are determined from the zeroth-order Wigner functions,
where $N_{0}=n,n_5,\epsilon$ corresponding to $i=2,2,3$ and $Z_{N0}=V_{0},A_{0},V_{0}$, respectively.
The pressure is given by $P=\epsilon /3$.
Coefficients $\xi$, $\xi_B$, $\xi_5$ and $\xi_{B5}$ are given by
\begin{equation}
\Xi = c \pi \int d p_0 p_0^j [\theta (p_0)-\theta (-p_0)] Z_{\Xi 0},
\end{equation}
where $\Xi = \xi, \xi_B, \xi_5, \xi_{B5}$ corresponding to
$j=1,0,1,0$, $c=2,Q,2,Q$, and $Z_{\Xi 0}=A_{0},A_{0},V_{0},V_{0}$, respectively.
It is easy
to verify following relations: $\xi=(1/2)\partial n_5/\partial \mu$, $\xi_5=(1/2)\partial n/\partial \mu$, 
$\xi_B = (Q/2)\partial\xi/\partial \mu$, and $\xi_{B5} = (Q/2) \partial\xi_5/\partial \mu$. 

One can complete the above integrals analytically to obtain   
coefficients $\xi$, $\xi _B$, $\xi _5$ and $\xi _{B5}$ of the induced currents
as functions of $\mu$, $\mu _5$ and $T$, 
\begin{eqnarray}
\label{xi-final}
 \xi &=& \frac{1}{\pi^2}\mu \mu_5 ,\,\,\,
\xi_B  = \frac{Q}{2\pi^2}\mu_5 ,\\
\label{xi5-final}
 \xi_5 &=&\frac{1}{6} T^2 + \frac{1}{2\pi^2}\left(\mu^2+\mu_5^2\right) , \,\,\,
\xi_{B5} = \frac{Q}{2\pi^2} \mu .
\end{eqnarray}
Thermodynamical quantities $n$, $n_5$ and $\epsilon $ can be similarly obtained.

The current in Eq.~(\ref{jjv}) induced by magnetic field and vorticity with coefficients $\xi_B$ and $\xi $ 
in Eq.~(\ref{xi-final}), known as the CME and CVE \cite{Son:2009tf,Fukushima:2008xe,Kharzeev:2010gr},
respectively, is a direct consequence of the quantum kinetic equation for the Wigner function.
The axial-vector current in Eq.~(\ref{jj5}) induced by magnetic field and vorticity corresponds 
to some sort of reversed CME and CVE, respectively. 
These results are consistent to those obtained from the second law
of thermodynamics in Refs.~\cite{Pu:2010as} and \cite{Sadofyev:2010pr}
except a quadratic term in temperature in $\xi _5$ induced by vorticity. 
It should be noted that Eqs.~(\ref{xi-final}-\ref{xi5-final}), including the temperature term in $\xi_5$, 
have also been obtained independently in Ref.~\cite{Landsteiner:2011cp} within the Kubo formalism. 

Conservation equations for $j^\mu$ and $j^\mu_5$
\begin{equation}
\label{vector-convervation}
\partial_\mu j^\mu =0 ,\,\,\, 
\partial_\mu j_5^\mu = -\frac{Q^2}{2\pi^2} E\cdot B,
\end{equation}
can be derived from Eqs.~(\ref{jjv}-\ref{jj5}) with constraints on fluid and thermodynamical 
variables in Eq.~(\ref{relation-static}). The electric field in the chiral anomaly appears 
through $\partial_\sigma \mu = - QE_\sigma$ from Eq.~(\ref{relation-static}).
Note that we derived the chiral anomaly here without regularization in contrast to the derivation in quantum field theory.
This is because the Wigner function contains two fermionic fields separated in 
space-time (nonlocal) and therefore free of singularities. 
One can also verify the energy-momentum conservation equation in the background field,
\begin{equation}
\label{emt1}
\partial _\mu T^{\mu\nu}=QF^{\nu\rho} j_\rho,
\end{equation}
from Eqs.~(\ref{jjv}) and (\ref{emt}) with constraints in Eq.~(\ref{relation-static}). It is interesting to observe that
constraints in Eq.~(\ref{relation1}) or (\ref{relation-static})
require $\omega ^\mu \parallel B^\mu \parallel E^\mu$, which is
crucial for the energy-momentum conservation in Eq.~(\ref{emt1}).

It is remarkable that we have derived from  the quantum kinetic equation not only currents in 
Eqs.~(\ref{jjv}-\ref{jj5}) with their coefficients in 
Eqs.~(\ref{xi-final}-\ref{xi5-final}) but also a complete set of conservation equations with the chiral anomaly in
Eqs.~(\ref{vector-convervation}-\ref{emt1}) for charge, chiral
charge and energy-momentum, respectively. In contrast, these
conservation equations are used as inputs to obtain the currents 
in Refs.~\cite{Son:2009tf,Pu:2010as} 
with the requirement of the second law of thermodynamics.  

{\it Multi-flavor fluid.} --- So far we have only considered a fluid with a single type of fermions.
An extension to the case of multi-flavor quarks is straightforward. We can
consider a three-flavor fluid with $u$, $d$ and $s$ quark  and
their anti-quarks. Note that each quark carries $N_c$ fundamental color charges.
For the induced electromagnetic and baryonic vector current $j^{\mu}$,
\begin{eqnarray}
\label{3-flavor}
&& \xi^{\mathrm{baryon}} = \frac{N_c}{\pi^2}\mu
\mu_5,\; \xi_B^{\mathrm{baryon}} = \frac{N_c}{6\pi^2}\mu_5 \sum_f
Q_f ,
\nonumber\\
&& \xi^{\mathrm{EM}} = \frac{N_c}{\pi^2}\mu \mu_5 \sum_f Q_f,\;
\xi_B^{\mathrm{EM}} = \frac{N_c}{2\pi^2}\mu_5 \sum_f Q^2_f .
\end{eqnarray}
For this  three-flavor quark matter we have $\sum_f Q_f=0$, and
$\xi_B^{\mathrm{baryon}}=\xi^{\mathrm{EM}}=0$. This implies that the CME (CVE)
dominates the electromagnetic (baryonic) current \cite{Kharzeev:2010gr}.
For the induced baryonic axial-vector current $j^{\mu}_{5}$,
\begin{eqnarray}
\label{j5-baryon} \xi_5&=& N_c \left[\frac{1}{6} T^2
+\frac{1}{2\pi^2}(\mu^2+\mu_5^2)\right] ,\nonumber \\
\xi_{B5}&=& \frac{N_c}{6\pi^2}\mu \sum_f Q_f =0 .
\end{eqnarray}
Therefore, magnetic fields cannot induce the axial-vector current
in a three-flavor quark matter, which can only be induced by vorticity.

\begin{figure}
\includegraphics[scale=0.4]{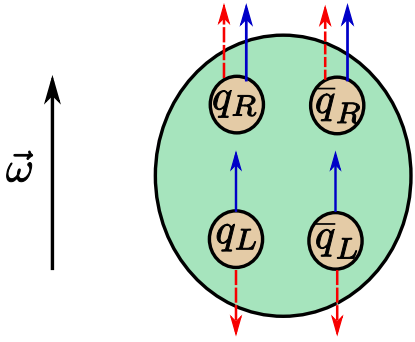}
\caption{\label{pol}(Color online) The axial current induced
by vorticity leads to the local polarization effect.
The momentum (spin) direction is in the red-dashed (blue-solid) arrow.}
\end{figure}

{\it Local polarization effect.} --- An axial-vector current induced by vorticity implies that
the right (left)-handed fermions move parallel (opposite) to the direction of vorticity. 
Since the momentum of a right (left)-handed massless fermion is parallel (opposite)
to its spin, all spins are parallel to the direction of vorticity (see Fig.~\ref{pol} for illustration). This results in 
the local polarization effect (LPE) similar to what was proposed in Refs.
\cite{Liang:2004ph,Becattini:2007sr,Gao:2007bc} due to spin-orbital coupling.
The LPE can be measured via hadron (e.g. hyperon) polarization along the direction of vorticity or the
global orbital angular momentum in non-central heavy-ion collisions \cite{Liang:2004ph}.
Note that $\xi_5$ in Eq.~(\ref{j5-baryon})  has three quadratic terms in $T$,  $\mu$ and $\mu_5$.
Therefore, the LPE should be present in both high and low energy heavy-ion collisions 
with either low baryonic chemical potential and high temperature or vice versa.


This work was supported by the NSFC
under grant Nos. 11105137, 11125524, 10975092 and 11221504, 
and by the U.S. DOE
under Contract No. DE-AC02-05CH11231 and within the framework of the
JET Collaboration. S.P. is supported in part by the China
Postdoctoral Science Foundation under the grant No. 2011M501046.

\end{document}